\documentclass[submission, Phys]{SciPost}

\usepackage{graphicx,amssymb,amsmath,empheq}
\usepackage{amsthm,authblk}
\usepackage{empheq}
\usepackage{subfig} 
\usepackage[title]{appendix}
\usepackage{hyperref}
\hypersetup{colorlinks = true, linkcolor=black, citecolor=black, urlcolor=blue}
\usepackage{url}
\usepackage{tikz,fp}
\usepackage{tikz-cd}
\usepackage{soul}
\usetikzlibrary{arrows}
\usetikzlibrary{intersections}
\usetikzlibrary{shapes.geometric}
\usetikzlibrary{decorations.pathmorphing, patterns,shapes,fixedpointarithmetic}
\usetikzlibrary{decorations.markings}

\pgfdeclarepatternformonly{south west lines}{\pgfqpoint{-0pt}{-0pt}}{\pgfqpoint{3pt}{3pt}}{\pgfqpoint{3pt}{3pt}}{
        \pgfsetlinewidth{0.4pt}
        \pgfpathmoveto{\pgfqpoint{0pt}{0pt}}
        \pgfpathlineto{\pgfqpoint{3pt}{3pt}}
        \pgfpathmoveto{\pgfqpoint{2.8pt}{-.2pt}}
        \pgfpathlineto{\pgfqpoint{3.2pt}{.2pt}}
        \pgfpathmoveto{\pgfqpoint{-.2pt}{2.8pt}}
        \pgfpathlineto{\pgfqpoint{.2pt}{3.2pt}}
        \pgfusepath{stroke}}

\tikzset{
  mid arrow/.style={postaction={decorate,decoration={
        markings,
        mark=at position .575 with {\arrow{stealth}}
      }}},
  near arrow/.style={postaction={decorate,decoration={
        markings,
        mark=at position .275 with {\arrow{stealth}}
      }}},
  far arrow/.style={postaction={decorate,decoration={
        markings,
        mark=at position .800 with {\arrow{stealth}}
      }}},
  snake arrow/.style={fixed point arithmetic, decorate, decoration={snake,amplitude=2pt, segment length=11pt},postaction={decoration={markings,mark=at position 0.625 with {\arrow{stealth}}},decorate}},
}

\begin{document}

\begin{center}{\Large \textbf{
Subsystem R\'enyi Entropy of Thermal Ensembles for SYK-like models
}}\end{center}

\begin{center}
Pengfei Zhang\textsuperscript{1*},
Chunxiao Liu\textsuperscript{2},
Xiao Chen\textsuperscript{3}
\end{center}

\begin{center}
{\bf 1} Institute for Quantum Information and Matter and Walter Burke Institute for Theoretical Physics, California Institute of Technology, Pasadena, California 91125, USA
\\
{\bf 2} Department of Physics, University of California, Santa Barbara, California 93106-9530, USA
\\
{\bf 3} Department of Physics, Boston College, Chestnut Hill, MA 02467, USA
\\
* PengfeiZhang.physics@gmail.com
\end{center}

\begin{center}
\today
\end{center}


\section*{Abstract}
{\bf
The Sachdev-Ye-Kitaev model is an $N$-modes fermionic model with infinite range random interactions. In this work, we  study the thermal R\'enyi entropy for a subsystem of the SYK model using the path-integral formalism in the large-$N$ limit. The results are consistent with exact diagonalization \cite{liu2018quantum} and can be well approximated by thermal entropy with an effective temperature \cite{huang2019eigenstate} when subsystem size $M\leq N/2$. We also consider generalizations of the SYK model with quadratic random hopping term or $U(1)$ charge conservation.}

\vspace{10pt}
\noindent\rule{\textwidth}{1pt}
\tableofcontents\thispagestyle{fancy}
\noindent\rule{\textwidth}{1pt}
\vspace{10pt}

\section{Introduction}
In these years, there is an increasing interest for the  entanglement/thermal entropy of many-body systems in both condensed matter and high-energy physics. On the condensed matter side, entanglement/thermal entropy can be used to identify the critical theory \cite{vidal2003entanglement,latorre2003ground,calabrese2004entanglement,Korepin_2004} as well as topological ordered phases \cite{kitaev2006topological,levin2006detecting}. It can be used to distinguish the many-body localization (MBL) phase \cite{Pal_2010,nandkishore2015many,abanin2019colloquium} and the thermal phase\cite{Deutsch_1991,Srednicki_1994}. In MBL phase, the entanglement entropy of the eigenstate has area law scaling\cite{Bauer_2013,Lukin256}, while in the thermal phase, the reduced density matrix of the typical state takes a thermal form and therefore the entanglement entropy obeys volume law scaling\cite{Garrison_2018}. Recent developments for the study of the information scrambling also focus on the entanglement entropy dynamics \cite{Calabrese_2009,lashkari2013towards,kitaev2014hidden,hosur2016chaos,fan2017out,Nahum_2017,Ho_2017,Mezei_2017}. As an interesting example, the projective measurement can induce a transition from the entanglement entropy perspective in systems with random unitary evolution \cite{li2018quantum,chan2019unitary,skinner2019measurement}. On the high energy side, the discovery of Ryu-Takayanagi formula \cite{Ryu:2006bv,Ryu:2006ef,Lewkowycz:2013nqa} directly relates the entanglement/thermal entropy of a subsystem to holographic bulk geometry, which has been latterly generalized to the time dependent case \cite{Hubeny:2007xt} or with quantum corrections \cite{Faulkner:2013ana,Engelhardt:2014gca}. Recent refinement of the Ryu-Takayanagi formula gives an plausible solution to the information paradox \cite{penington2019entanglement,almheiri2019entropy,almheiri2019page,almheiri2019islands,almheiri2019replica,penington2019replica}.

However, for general interacting many-body system without holographic description or conformal symmetry, there is no efficient method to study the behavior of entropy, especially for strongly correlated systems. Fortunately, a solvable model describing $N$ Majorana modes with infinite range random interaction has been proposed \cite{kitaev2014hidden,maldacena2016remarks}, which is related to the early work of Sachdev and Ye \cite{sachdev1993gapless}. This model is now known as the Sachdev-Ye-Kitaev (SYK) model. The studies show that the SYK model is a non-Fermi liquid without quasiparticles and has emergent conformal symmetry  at low energies \cite{kitaev2014hidden,maldacena2016remarks}. The low energy Schwarzian action \cite{kitaev2014hidden,maldacena2016remarks,kitaev2018soft} also matches the result of NAdS$_2$ dilaton gravity \cite{kitaev2018soft,maldacena2016conformal}. Latterly, different generalizations have been proposed, including a complex fermion version \cite{davison2017thermoelectric,gu2019notes} and the SYK model with extra random hopping terms \cite{banerjee2017solvable,chen2017competition,song2017strongly}, which have important applications in different fields. The hallmark of these models is that the computation of correlation functions is reduced to integral equations in the large-N limit. This also leads to the ability to compute quantum dynamics precisely \cite{eberlein2017quantum,haldar2019quench,kuhlenkamp2019periodically,zhang2019evaporation,almheiri2019universal}.

In this work, we study the SYK model and its generalizations from the subsystem entanglement/thermal entropy perspective. We use the path integral method and derive the saddle point solution of the second R\'enyi entropy by taking the large $N$ limit. We solve the saddle point equation numerically and obtain the scaling of the second R\'enyi entropy in various cases. In particular, we compare the results at extremely low temperature with the entanglement entropy result of SYK model studied by using exact diagonalization for finite size systems \cite{liu2018quantum} and using the subsystem thermalization arguments \cite{huang2019eigenstate}\footnote{There are also studies of entropy dynamics for coupled SYK models prepared in an thermofield double state \cite{penington2019replica,gu2017spread,Yiming}, which is related to the black hole evaporation problem. }. We find the results are consistent with the subsystem size is smaller than half of the system size. 

The rest of the paper is organized as follows: In Section \ref{sec:review}, we  briefly review the SYK model and the subsystem thermalization arguments. In Section \ref{sec:path_integral}, we recall the path-integral formalism \cite{liu2018quantum} for the R\'enyi entropy, where the action is written in terms of bi-local fields. In the large-N limit, we derive the saddle point solution of the R\'enyi entropy and analytically compute it in various limits. In Section \ref{sec:num}, we solve the saddle point solution and present the numerical results for the second R\'enyi entropy of SYK models with different subsystem size and temperature. We find the argument in Section \ref{sec:review} works well despite small deviation from the analytical approximation \cite{huang2019eigenstate}. We further study the generalizations of SYK model by introducing random hopping quadratic term or considering the complex SYK model with charge conservation law. We conclude our results in the Section \ref{sec:conclusion}.

\section{Review of the SYK model and its entanglement}
\label{sec:review}
The SYK$_q$ model describes $q$-body random interacting Majorana fermions, the Hamiltonian is written as:
\begin{equation}
H=\frac{1}{q!}\sum_{i_1i_2...i_q}i^{q/2}J_{i_1i_2...i_q}\chi_{i_1}\chi_{i_2}...\chi_{i_q}.\label{H0}
\end{equation}
Here $q$ is an even integer. $i=1,2...N$ labels different Majorana modes with commutation relation $\{\chi_i,\chi_j\}=\delta_{ij}$. The $J_{i_1i_2...i_q}$ with different indices are independent Gaussian variables with zero means. Their variance is given as:
\begin{equation}
\overline{|J_{i_1i_2...i_q}|^2}=\frac{(q-1))!J^2}{N^{q-1}}. \label{Var}
\end{equation}

In the large-$N$ limit, the two-point function in thermal equilibrium $G_\beta(\tau)=\left<\mathcal{T}\chi_i(\tau)\chi_i(0)\right>$ is computed by taking melon diagrams \cite{kitaev2014hidden,maldacena2016remarks}. The Swinger-Dyson equation is written as 
\begin{equation}
G^{-1}_\beta(i\omega_n)=-i\omega_n-\Sigma_\beta(i\omega_n),\ \ \ \ \ \ \Sigma_\beta(\tau)=\begin{tikzpicture}[baseline={([yshift=-6pt]current bounding box.center)}, scale=1.2]
\draw[thick] (-24pt,0pt) -- (-15pt,0pt);
\draw[thick,dashed] (-15pt,0pt)..controls (-8pt,18pt) and (8pt,18pt)..(15pt,0pt);
\draw[thick] (-15pt,0pt)..controls (-8pt,10pt) and (8pt,10pt)..(15pt,0pt);
\draw[thick] (-15pt,0pt)..controls (-8pt,-10pt) and (8pt,-10pt)..(15pt,0pt);
\draw[thick] (15pt,0pt) -- (-15pt,0pt);
\draw[thick] (15pt,0pt) -- (24pt,0pt);
\end{tikzpicture}=J^2G_\beta^{q-1}(\tau).\label{schwingerdyson}
\end{equation}
Here $\omega_n=2\pi(n+1/2)/\beta$ is the Matsubara frequency. In the low energy limit, by neglecting the $i\omega_n$ term, the Green's function (at $J^{-1}\ll\tau\ll\beta$) is found to be proportional to $\text{sgn}(\tau)/|\tau|^{2\Delta}$ with $\Delta=1/q$ \cite{maldacena2016remarks}. Here $\Delta$ is identified with the scaling dimension of fermions. As a result, the spectral function for small frequency diverges as $\omega^{2\Delta-1}$ for $q\geq 4$. This is understood as a non-Fermi liquid behavior. For $q=2$, the model is a quadratic random hopping model, and the single particle spectral shows a semi-circle law with finite weight as $\omega \rightarrow 0$.

The free energy, or the on-shell action, is determined by the solution of \eqref{schwingerdyson}  \cite{maldacena2016remarks}:
\begin{equation}
I_\beta=\beta \mathcal{F}=-N\sum_n\log(-i\omega_n-\Sigma_\beta(i\omega_n))+\frac{\beta N}{2}\int d\tau \left(\Sigma_\beta(\tau)G_\beta(\tau)-\frac{J^2G_\beta^q(\tau)}{q}\right).
\end{equation}
The thermodynamical entropy $\mathcal{S}$ and the energy $\mathcal{E}$ are then determined by taking derivatives. It is known that for $q\geq 4$, there is extensive entropy at zero temperature limit  \cite{kitaev2014hidden,maldacena2016remarks}:
\begin{equation}
s(\Delta)\equiv\frac{\mathcal{S}(\Delta)}{N}=\int_\Delta^{1/2} dx\  \pi  \left(\frac{1}{2}-x\right) \tan (\pi  x)\label{Sthe}
\end{equation}
which reflects the existence of exponential number of low-energy states in many-body spectral. We have $\mathcal{S}(1/2)=0$, $\mathcal{S}(1/4)\approx0.2324N$ and $\mathcal{S}(0)=\log(2)N/2$.

Now let's review the argument \cite{huang2019eigenstate} for computing the subsystem entropy of the SYK model. We consider the full system is prepared in a state described by a density matrix $\rho$. Here $\rho$ can be either a pure state or an ensemble. To define the entropy of a subsystem, we divide the system into $A$ and $B$, where $A$ contains $M$ Majorana fermions and $B$ contains $N-M$ Majorana fermions. We take the large-$N$ limit with fixed ratio $\lambda=M/N$. The $n$-th R\'enyi entropy of the subsystem $A$ is given by 
\begin{equation}
\mathcal{S}_A^{(n)}=\frac{1}{1-n}\log(\text{tr}_A\rho_A^n),\ \ \ \ \ \ \rho_A=\text{tr}_B\rho.
\end{equation}
The Von Neumann entropy is given by taking $n\rightarrow 1$.

The energy $\mathcal{E}$ of the total system is given by $\mathcal{E}=\text{tr}(\rho H)$. For the subsystem $A$, its Hamiltonian $H_A$ is defined as restricting $H$ into the subsystem $A$:
\begin{equation}
H_A=\frac{1}{q!}\sum_{i_1i_2...i_q\in A}J_{i_1i_2...i_q}\chi_{i_1}\chi_{i_2}...\chi_{i_q}. \label{HA}
\end{equation} 
This is again an SYK model, but with effective $\tilde{J}=\lambda^{\frac{q-1}{2}}J$ because of the $N$ dependence in \eqref{Var}. If we assume the $\rho$ is symmetric under the permutation of indices. The energy of the subsystem is given by $\mathcal{E}_A=\frac{M^q}{N^q}\mathcal{E}$. The energy density in $A$ is then $\mathcal{E}_A/M=\lambda^{q-1}\mathcal{E}/N$. Using $\mathcal{E}^0$ to denote the ground state energy of corresponding system, we have:
\begin{equation}
\epsilon_A\equiv\frac{\mathcal{E}_A}{\mathcal{E}^0_A}=\lambda^{\frac{q-1}{2}}\frac{\mathcal{E}}{\mathcal{E}^0}, \ \ \ \ \ \ \mathcal{E}^0_A=\lambda^{\frac{q-1}{2}}\mathcal{E}^0.\label{energydensity}
\end{equation} 
Assuming the system thermalizes \cite{hunter2018thermalization,haque2019eigenstate,sonner2017eigenstate}, the proposal \cite{huang2019eigenstate} is to approximate $\rho_A$ by a thermal ensemble with with energy $\mathcal{E}_A$. For pure states, this approximation can only work if $\lambda<1/2$. Then we have $s_A^{(1)}=\mathcal{S}_A^{(1)}/M=s(\Delta,\epsilon_A)$\footnote{Here if we consider general $n\neq 1$, the result would depend on whether we assume the ensemble to be canonical or micro-canonical. Moreover, for local systems, studies show that equation \eqref{energydensity} should be modified \cite{huang2019universal,vidmar2017entanglement,lu2019renyi,murthy2019structure}. However, here we have checked that for non-local system such as SYK-like models, the correction is small. This is consistent with the Figure \ref{fig1}. }, here $s(\Delta,\epsilon_A)$ is the entropy density at relative energy density $\epsilon_A$ in the thermodynamic limit. As an approximation which is good away from $\epsilon_A=1$ \cite{garcia2017analytical}:
\begin{equation}
s(\Delta,\epsilon_A)=\frac{\log(2)}{2}-\frac{1}{q^2}\arcsin^2(\epsilon_A).\label{ana}
\end{equation}
In the limit of $\lambda\rightarrow0$, the energy density becomes zero, which corresponds to a maximal mixed state with $S_A^{(n)}=M \log(2)/2$. This works for any $\lambda<1$ if we take $q\rightarrow \infty$.

\section{Path-integral formulation for the R\'enyi entropy}
\label{sec:path_integral}
\subsection{The formulation}
In this section, we consider the path-integral representation for computing the subsystem R\'enyi entropy \cite{liu2018quantum}, restricted to a thermal ensemble $\rho=e^{-\beta H}/Z$ of the total system. Here $Z=\text{tr}e^{-\beta H}=\exp(-I_\beta)$.

We firstly consider the reduced density matrix $\rho_A$. A pictorial representation of the density matrix $\rho$ is given by:
\begin{equation}
\rho=\frac{1}{Z}\times
\begin{tikzpicture}[baseline={([yshift=0pt]current bounding box.center)}]
   \draw (1,-0.2) arc(0:-180:1 and 1);
   \draw (1.3,-0.2) arc(0:-180:1.3 and 1.3);

       \draw[dotted,thick] (0.866,-0.725) -- (1.126,-0.875);

        \draw[dotted,thick] (0.5,-1.091) -- (0.65,-1.35);

              \draw[dotted,thick] (0,-1.225) -- (0,-1.525);

        \draw[dotted,thick] (-0.5,-1.091) -- (-0.65,-1.35);

           \draw[dotted,thick] (-0.866,-0.725) -- (-1.126,-0.875);

      \draw (0.4,-0.8) node{$B$};
       \draw (0.9,-1.4) node{$A$};
       \draw (-1.5,-0.2) node{$\beta$};
       \draw (1.5,-0.2) node{$0$};
\end{tikzpicture}.
\end{equation}
Here the free end of the solid line denotes the possible quantum states. Imposing the boundary condition on the end gives the matrix element of density matrix. The dashed line represents the interaction between subsystems. To find the reduced density matrix for system $A$, we trace out the subsystem $B$:
\begin{equation}
\rho_A=\text{tr}_B\rho=\frac{1}{Z}\times
\begin{tikzpicture}[baseline={([yshift=0pt]current bounding box.center)}]
   \draw (0.65,-0.65) arc(0:-360:0.65 and 0.65);
   \draw (1.3,-0.2) arc(0:-180:1.3 and 1.3);

       \draw[dotted,thick] (0.6,-0.4) -- (1.126,-0.875);

        \draw[dotted,thick] (0.5,-1.091) -- (0.65,-1.35);

              \draw[dotted,thick] (0,-1.35) -- (0,-1.475);

        \draw[dotted,thick] (-0.5,-1.091) -- (-0.65,-1.35);

           \draw[dotted,thick] (-0.6,-0.4) -- (-1.126,-0.875);

      \draw (0.,-0.6) node{$B$};
       \draw (0.9,-1.4) node{$A$};
       \draw (-1.5,-0.2) node{$\beta$};
       \draw (1.5,-0.2) node{$0$};
       \filldraw  (0pt,0.15pt) circle (1.5pt) node[left]{\scriptsize $ $};
\end{tikzpicture}.
\end{equation}
Here the black dot represents the anti-periodic boundary condition for fermionic fields \cite{altland2010condensed}. Then $\text{tr}_A \rho_A^n$ can be computed by sewing $n$ reduced density matrix with an overall anti-periodic boundary condition. To be concrete, from now on we focus on the $n=2$ case, which gives a lower bound of the Von Neumann entropy. We have:
\begin{equation}
\exp(-\mathcal{S}_A^{(2)})=\text{tr}_A \rho_A^2=\frac{1}{Z^2}\times \left[\begin{tikzpicture}[baseline={([yshift=0pt]current bounding box.center)}]
   \draw (0.6,-0.85) arc(0:-360:0.6 and 0.6);
   \draw (0.6,0.45) arc(0:-360:0.6 and 0.6);
   \draw (1.3,-0.2) arc(0:-360:1.3 and 1.3);

        \draw[dotted,thick] (0.4,-0.4) -- (1.126,-0.875);

        \draw[dotted,thick] (0.5,-1.191) -- (0.65,-1.35);

        \draw[dotted,thick] (0,-1.44) -- (0,-1.475);

        \draw[dotted,thick] (-0.5,-1.191) -- (-0.65,-1.35);

        \draw[dotted,thick] (-0.4,-0.4) -- (-1.126,-0.875);

        \draw[dotted,thick] (0.4,0) -- (1.126,0.475);

        \draw[dotted,thick] (0.5,0.791) -- (0.65,0.95);

        \draw[dotted,thick] (0,1.04) -- (0,1.075);

        \draw[dotted,thick] (-0.5,0.791) -- (-0.65,0.95);

        \draw[dotted,thick] (-0.4,0) -- (-1.126,0.475);

      \draw (0.,-0.85) node{$B_1$};
      \draw (0.,0.45) node{$B_2$};
       \draw (0.9,-1.4) node{$A$};
       \draw (-1.5,-0.2) node{$\beta$};
       \draw (1.5,0.2) node{$2\beta$};
       \draw (1.5,-0.6) node{$0$};
       \filldraw  (0pt,-7pt) circle (1pt) node[left]{\scriptsize $ $};
       \filldraw  (0pt,-4pt) circle (1pt) node[left]{\scriptsize $ $};
       \filldraw  (37pt,-5.5pt) circle (1pt) node[left]{\scriptsize $ $};
\end{tikzpicture}\right]=\exp(-I^{(2)}+2I_\beta).
\end{equation}
We then have $\mathcal{S}_A^{(2)}=I^{(2)}-2I_\beta$. The generalization to the $n$-th R\'enyi entropy is straightforward. More explicitly, we have
\begin{equation}
\begin{aligned}
e^{-I^{(2)}}&=\int_{\text{b.c.}}\mathcal{D}\chi_{i}(\tau) \exp(-S^{(2)}[\chi_{i}]).\\
S^{(2)}&= \int_0^{2\beta}d\tau\left(\frac{1}{2}\sum_i\chi_{i}\partial_\tau\chi_{i}+\frac{1}{q!}\sum_{i_1i_2...i_q}i^{q/2}J_{i_1i_2...i_q}\chi_{i_1}\chi_{i_2}...\chi_{i_q}\right).
\end{aligned}
\end{equation}
Here the boundary condition is given by
\begin{equation}
\begin{aligned}
&\chi_{i}(0^+)=-\chi_{i}(2\beta^-),\ \ \ \ \ \chi_{i}(\beta^+)=\chi_{i}(\beta^-),\ \ \ \ \ \ \text{if} \ \ \ i\in A;\\
&\chi_{i}(0^+)=-\chi_{i}(\beta^-),\ \ \ \ \ \chi_{i}(\beta^+)=-\chi_{i}(2\beta^-),\ \ \ \ \text{if} \ \ \ i\in B.\label{bc}
\end{aligned}
\end{equation}
Note that these boundary conditions break the time translation symmetry of the system explicitly. 

To proceed, we should average over the disorder configuration of $J_{i_1i_2...i_q}$ as $\overline{\mathcal{S}_A^{(2)}}=-\overline{\log \text{tr}_A \rho_A^2}$. However, this computation needs to introduce additional disorder replicas. In the large-$N$ limit, both numerical analysis and analytical arguments suggest the validity of disorder replica diagonal ansatz for the SYK model to the leading order of $1/N$ expansion \cite{fu2016numerical,gur2018does,kitaev2018soft,gu2019notes}.  We assume that similar arguments apply here and make the approximation:
\begin{equation}
\overline{\mathcal{S}_A^{(2)}}=-\overline{\log \text{tr}_A \rho_A^2}\approx -\log\left[\frac{\overline{e^{-I^{(2)}}}}{\overline{Z}^2}\right].
\end{equation}
It is then straightforward to integrate out $J_{i_1i_2...i_q}$ and introduce the $G-\Sigma$ action according to the standard procedure \cite{maldacena2016remarks}. The main difference is that we should introduce two sets of bi-local fields, in order to take different boundary conditions into account. Explicitly, we define 
\begin{equation}
G_A(\tau,\tau')=\frac{1}{M}\sum_{i\in A}\chi_i(\tau)\chi_i(\tau'),\ \ \ \ \ \ G_B(\tau,\tau')=\frac{1}{N-M}\sum_{i\in B}\chi_i(\tau)\chi_i(\tau').
\end{equation}
Additional Lagrangian multiplier is introduced into the path integral as:
\begin{equation}
\begin{aligned}
\delta\left(G_A-\frac{1}{M}\sum_{i\in A}\chi_i\chi_i\right)&=\int \mathcal{D}\Sigma_A\ e^{\frac{1}{2}\int d\tau d\tau'\Sigma_A(\tau,\tau')\left(\sum_{i\in A}\chi_i(\tau)\chi_i(\tau')-MG_A(\tau,\tau')\right)},\\
\delta\left(G_B-\frac{1}{N-M}\sum_{i\in B}\chi_i\chi_i\right)&=\int \mathcal{D}\Sigma_B\ e^{\frac{1}{2}\int d\tau d\tau'\Sigma_B(\tau,\tau')\left(\sum_{i\in B}\chi_i(\tau)\chi_i(\tau')-(N-M)G_B(\tau,\tau')\right)}.
\end{aligned}
\end{equation} 
After further integrating out the Majorana field, one obtains the $G-\Sigma$ action for computing the subsystem entropy:
\begin{equation}
\overline{e^{-I^{(2)}}}=\int \mathcal{D}G_A \mathcal{D}G_B \mathcal{D}\Sigma_A \mathcal{D}\Sigma_B \exp(-S^{(2)}),
\end{equation}
with an action 
\begin{equation}
\begin{aligned}
S^{(2)}=&-\frac{M}{2}\log \underset{A}{\det} (\partial_\tau-\Sigma_A)-\frac{N-M}{2}\log \underset{B}{\det} (\partial_\tau-\Sigma_B) \\
&+\frac{M}{2}\int d\tau d\tau'G_A(\tau,\tau')\Sigma_A(\tau,\tau')+\frac{N-M}{2}\int d\tau d\tau'G_B(\tau,\tau')\Sigma_B(\tau,\tau')\\
&-\frac{J^2}{2qN^3}\int d\tau d\tau' \left(MG_A(\tau,\tau')+(N-M)G_B(\tau,\tau')\right)^q. \label{action}
\end{aligned}
\end{equation}
Here the label under the $\det$ indicates the boundary condition \eqref{bc}. The action is proportional to $N$. Consequently, we could use a saddle point approximation to the leading order of $1/N$. The saddle point equation is 
\begin{equation}
G_A=(\partial_\tau-\Sigma_A)^{-1}_A,\ \ \ \ \ \ G_B=(\partial_\tau-\Sigma_B)^{-1}_B,\ \ \ \ \ \ \Sigma_A=\Sigma_B=J^2(\lambda G_A+(1-\lambda)G_B)^{q-1}. \label{SD}
\end{equation}
Diagrammatically, this again corresponds to melon diagrams, while internal lines can be either in $A$ or $B$, with the corresponding weight. Note that although the self-energy is the same for both $A$ and $B$, the inverse for computing $G_A$ and $G_B$ is under different boundary condition, which leads to different results. After solving these equations self-consistently, we can then compute the action \eqref{action} and obtain the entropy $\mathcal{S}_A^{(2)}$. 

Finally, although we use the original SYK model as an example, it is straightforward to generalize the derivation to any SYK-like models with an effective $G-\Sigma$ action\footnote{For tensor models, the two point Green's function satisfy similar equations. However, the action could not be computed easily. }. We will present two examples in the next section. 

\subsection{Analysis in different limits}

For $\lambda\rightarrow1$, all Majorana modes are in $B$. Consequently, the path-integral reduces to a thermal ensemble with inverse temperature $2\beta$ and the final result $\mathcal{S}_A^{(2)}$ reproduces the thermal R\'enyi entropy of the full system.

A more non-trivial limit is $\lambda \rightarrow 0$. For $\lambda=0$, the contribution from $I^{(2)}$ and $I_\beta$ cancels, leading to $S^{(2)}_A=0$ as expected. To get the leading order contribution for small $\lambda$, we need to expand the action \eqref{action} around $\lambda=0$ and then computing the action using the $\lambda=0$ solution of Green's functions \eqref{SD}. The action to the leading order of $\lambda$ is written as:
\begin{equation}
\begin{aligned}
\delta S^{(2)}&=-\frac{M}{2}\log \underset{A}{\det} (\partial_\tau-\Sigma_A)+\frac{M}{2}\log \underset{B}{\det} (\partial_\tau-\Sigma_B)\\&=\frac{M}{2}\log \underset{A}{\det} (G_A)-\frac{M}{2}\log \underset{B}{\det} (G_B). \label{deltaS}
\end{aligned}
\end{equation}
Here we have used the Schwinger-Dyson equation \eqref{SD} at $\lambda=0$. To proceed, we need  to know the specific form of the Green's functions. For $G_B$, at $\lambda=0$, the solution is block diagonal:
\begin{equation}
G_B(\tau,\tau')=G_\beta(\tau-\tau')\left\{\theta(\beta-\tau)\theta(\beta-\tau')+\theta(\tau-\beta)\theta(\tau'-\beta)\right\},
\end{equation}
while for $G_A$, the boundary condition gives off-diagonal components:
\begin{equation}
G_A(\tau,\tau')=G_B(\tau,\tau')+2G_\beta(\tau)G_\beta(\tau')\left\{\theta(\tau-\beta)\theta(\beta-\tau')-\theta(\beta-\tau)\theta(\tau'-\beta)\right\}.
\end{equation}
Here the additional term imposes the boundary condition \eqref{bc}. We then need to compute \eqref{deltaS} using these solutions of Green's functions. The important observation is that: 

1. The Green's function $G_B$ is of the same form  as two copies of $G_\beta$, which is the Green's function on a thermal ensemble. 

2. The Green's function $G_A$ is of the same form as that of the Kourkoulou-Maldacena states \cite{kourkoulou2017pure}. Consider pairing the Majorana fields as $c_j=\frac{\chi_{2j-1}+i\chi_{2j}}{2}$, the Kourkoulou-Maldacena states is given by applying imaginary evolution to the eigenstate of $n_j=c_j^\dagger c_j$:
\begin{equation}
|KM\rangle=e^{-\beta H/2}|s\rangle,\ \ \ \ \ \ n_j|s\rangle=s_j|s\rangle, \ \ \ \text{with}\ \ \ s_j=0\ \text{or}\ 1.
\end{equation}

As a result, $e^{-\delta S^{(2)}}$ corresponds to the difference between $\left<KM|KM\right>=\langle s| e^{-\beta H}|s\rangle$ and $\text{tr}e^{-\beta H}$ with Majorana fermion number $2M$. However, we know that for different ${s_j}$, $\left<KM|KM\right>$ should be the same, and proportional to corresponding $\text{tr}e^{-\beta H}$ with a possible factor that does not depend on the interaction strength \cite{kourkoulou2017pure}. We could just determine the factor by considering $J \rightarrow 0$ \footnote{Note that in the path integral formulation of $\left<KM|KM\right>$, there may be an additional factor absorbed in measure $\mathcal{D}\chi$. Here we determine the factor by using some known limits.}. This leads to $S_A^{(n)}=M \log(2)/2$ for small $\lambda$ and any $\beta J$, consistent with analysis in the previous section.

\section{Numerical results for the SYK-like models}
\label{sec:num}
Equations \eqref{action} and \eqref{SD} give a complete answer to the subsystem entropy in the large-$N$ limit. However, for general $\lambda$ it is hard to study them analytically due to the lack of time translational invariance. Instead here we perform a numerical study. In this section, we first describe our numerical procedure, and then presents numerical results for different SYK-like models. 

Numerically, we need to first solve \eqref{SD} iteratively. We discretize the imaginary time $\tau\in[0,2\beta)$ to $L$ points, then both $G_s$ and $\Sigma_s$ (for $s=A, B$) become $L\times L$ matrices. We typically take $L\sim 200\text{--}300$. After the discretization, the Schwinger-Dyson equation is written as:
\begin{equation}
(G_s)_{ij}=\left((G_s^0)^{-1}-\Sigma_s\right)^{-1}_{ij},\ \ \ \ \ \ (\Sigma_s)_{ij}=J^2\left(\frac{2\beta}{L}\right)^2\left(\lambda (G_A)_{ij}+(1-\lambda) (G_B)_{ij} \right)^{q-1}.
\end{equation}
Here, the boundary condition \eqref{bc} is taken into account by the bare Green's function $G^0_s$. Explicitly, we have \cite{Yiming}: 
\begin{equation}
\begin{aligned}
(G_A^0)_{ij}&=\frac{1}{2}\text{sgn}(i-j),\ \ \ \ \ \ \text{for}\ i,j\in\{1,2...L\},\\
(G_B^0)_{ij}&=\frac{1}{2}\text{sgn}(i-j),\ \ \ \ \ \ \text{for}\ i,j\in\{1,2...L/2\}\ \text{or}\ \in\{L/2+1,L/2+2...L\},
\end{aligned}
\end{equation}
and zero otherwise. Here $\text{sgn}(x)$ is the sign function with $\text{sgn}(0)=0$. Importantly, to obtain accurate numerical result with small number of points, we discretize $G^0_s$ and take the inverse to get $(G^0_s)^{-1}$, instead of using the discretized version of $\partial_\tau$, which would lead to an oscillating Green's function. 

After solving \eqref{SD}, we plug the results into \eqref{action} to compute the action. To ensure the convergence, we add $$0=\frac{M}{2}\left(\log \underset{A}{\det} (\partial_\tau)-\log 2\right)-\frac{N-M}{2}\left(\log \underset{B}{\det} (\partial_\tau)-2\log 2\right)$$ to the action. An extrapolation to $L\rightarrow \infty$ is performed finally by evaluating the action for different $L$.

\subsection{The original SYK model}

\begin{figure}[t]
  \center
  \includegraphics[width=0.75\columnwidth]{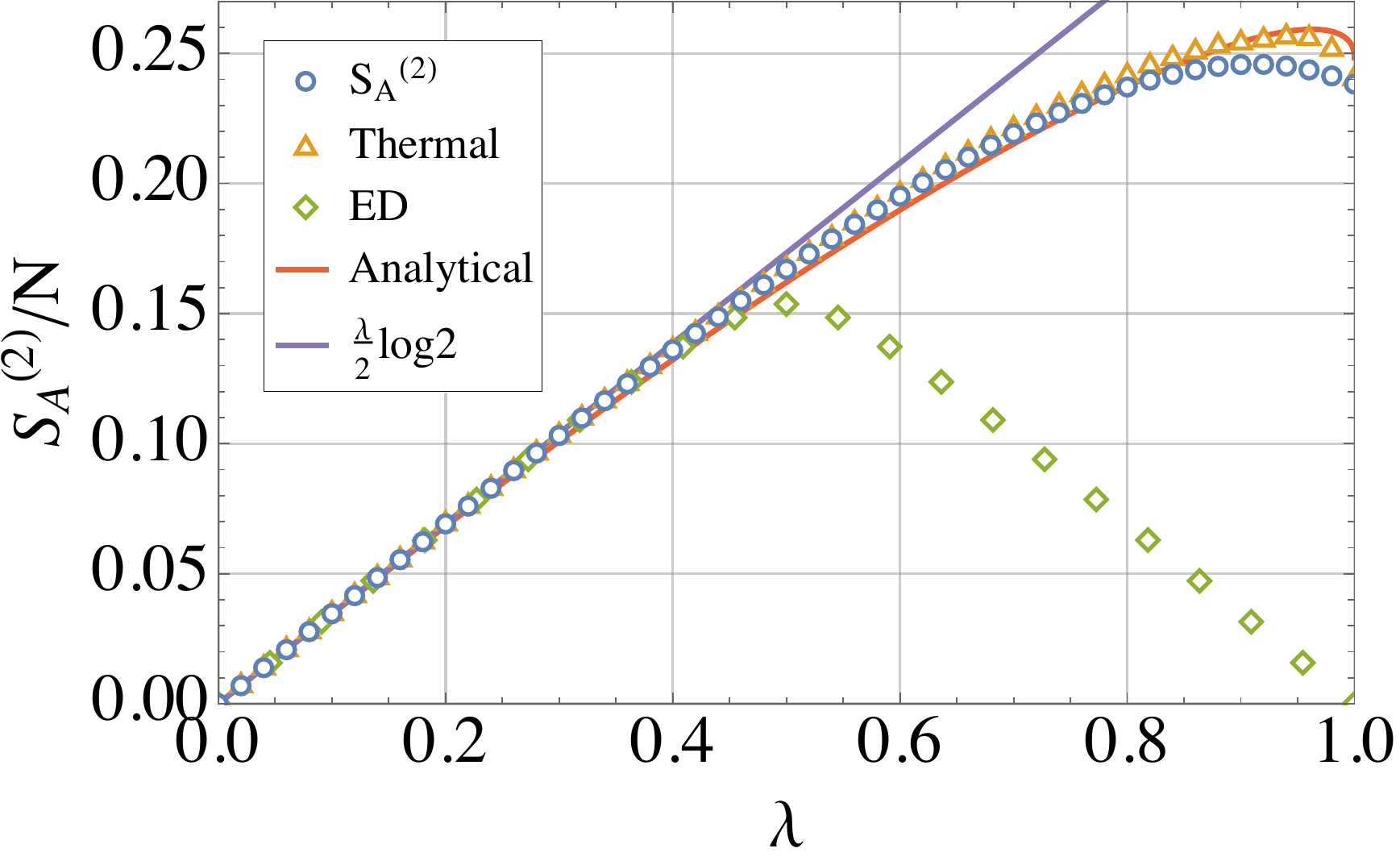}
  \caption{Subsystem entropy $\mathcal{S}_A^{(2)}/N$ for $\beta J=50$ and $q=4$ for different $\lambda$ (blue circle). The green square is the exact diagonalization result \cite{liu2018quantum}. The yellow triangle is thermal entropy with the effective temperature obtained by numerically matching the energy density \eqref{energydensity}. We also plot the analytical approximation \eqref{ana} using a red line and $\lambda\log 2/2$ using a purple line. }\label{fig1}
 \end{figure}
We firstly present our numerical results for the original SYK model. In Figure \ref{fig1}, we show a typical result for $\beta J=50$ with $q=4$ (blue circle). We take large $\beta J$ so that it corresponds to extremely low temperature. The green squares are results of the ground state entanglement entropy from exact diagonalization with $N=44$ Majorana fermions \cite{liu2018quantum}. We also compare $\mathcal{S}_A^{(2)}$ with the thermal R\'enyi entropy for the canonical ensemble (yellow circle), where the effective temperature of the corresponding ensemble is determined by matching the energy density \eqref{energydensity} numerically. The analytical approximation \eqref{ana} is shown in red line. 

\begin{figure}[t]
  \center
  \includegraphics[width=1\columnwidth]{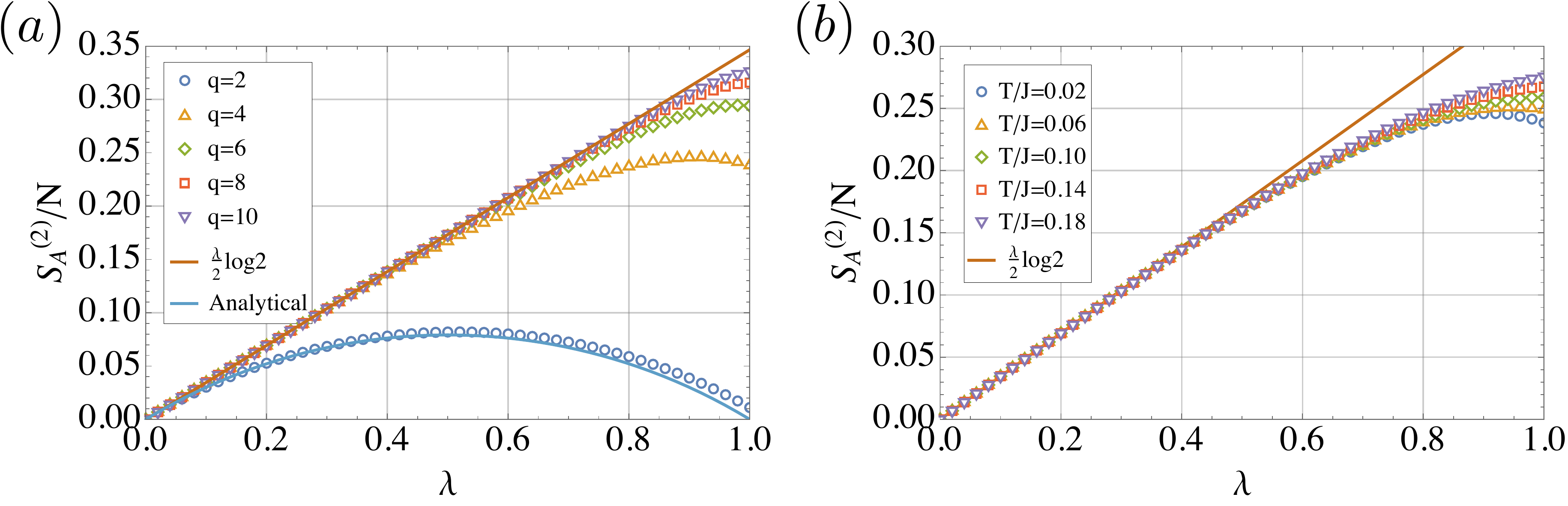}
  \caption{(a). Subsystem entropy $\mathcal{S}_A^{(2)}/N$ for different $q$. We fix $\beta \mathcal{J}=50/\sqrt{2}$, which corresponds to $\beta J=50$ for $q=4$. We also show the analytical result for the SYK$_2$ model, as computed in the appendix. (b). Subsystem entropy $\mathcal{S}_A^{(2)}/N$ for different temperature $T/J$ with $q=4$. In both figures we have also plotted $\lambda \log 2/2$ for reference. }\label{fig2}
 \end{figure}

We find that for $\lambda\lesssim0.4$, the numerical result for $\mathcal{S}_A^{(2)}$ matches the exact diagonalization result well. When $\lambda=1/2$, we find $\mathcal{S}_A^{(2)}(1/2)=0.334 N/2$. When the full system is prepared in a thermal ensemble, we find the subsystem entropy matches the thermal entropy well, except very close to $\lambda=1$. The analytical formula \eqref{ana} also gives a very good approximation, especially for small $\lambda \lesssim 0.4$.

We then study the dependence of subsystem entropy on $q$ and temperature $T$. We firstly consider the $q$ dependence. Here we use the standard convention \cite{maldacena2016remarks} and we fix the $\beta \mathcal{J}=50/\sqrt{2}$, where $\mathcal{J}$ is related to $J$ by $J^2=2^{q-1}\mathcal{J}^2/q$. This corresponds to $\beta J=50$ for $q=4$. According to \eqref{Sthe}, the thermal entropy even at zero temperature would approach $N\log 2/2$ in the large-$q$ limit. Consequently, the full curve for $\mathcal{S}_A^{(2)}$ will approach $\mathcal{S}_A^{(2)}=M\log2/2$ as $q\rightarrow \infty$, as shown in Figure \ref{fig2} (a). Similarly, as we increase the temperature, $\mathcal{S}_A^{(2)}$ will approach the same upper bound (See Figure \ref{fig2} (b)). For small $q$ or low temperature, $\mathcal{S}_A^{(2)}=M\log2/2$ is non-monotonic. In particular, when $q=2$, the zero temperature entropy density is zero and $\mathcal{S}_A^{(2)}$ in the limit $\beta\to \infty$ is fully symmetric around $\lambda=1/2$. The numerical result is shown in Figure \ref{fig2} (a) and is consistent with an analytical result derived from the random matrix theory\cite{liu2018quantum} (More details of this method is given in the appendix). Note that for all cases, the slope for $\mathcal{S}_A^{(2)}$ at small $\lambda$ collapses to $\lambda\log 2/2$, as proved in the previous section.

Before we end this subsection, we briefly discuss the connection between the entropy computed from the path integral method in the limit $\beta\to \infty$ and the entanglement entropy of the ground state. The path integral formulation introduced in the previous section is a standard method to compute the entropy for a subsystem. In systems with local interaction, at finite $\beta$, the  entropy for a subsystem is dominated by the thermal entropy and therefore is extensive. As we increase $\beta$ to $\infty$, the thermal entropy density decreases to zero and the  entropy is mainly coming from the entanglement entropy between the subsystem and its complement. The path integral formulation provides one method to compute the ground state entanglement entropy in the limit $\beta\to \infty$. However, the above physics needs to be modified in SYK$_4$ model since it has extensive zero temperature entropy. Therefore the zero temperature entropy of the subsystem and the entanglement between it and its complement can have same order and both of them contribute to the entropy computed from the path integral formulation. Since the zero temperature entropy is proportional to the subsystem size, the entropy is not symmetric at $\lambda=1/2$. Only when $\lambda<1/2$ is it roughly equal to entanglement entropy of the ground state as presented in Fig. \ref{fig1}. Notice that this  symmetry is fully restored in the SYK$_2$ model since the zero temperature entropy is zero.

\subsection{SYK model with random hopping}
In this subsection, we consider a specific generalization of the SYK model, which is the hybridization of SYK$_2$ and SYK$_q$ with $q\geq4$ \cite{banerjee2017solvable,chen2017competition,song2017strongly}:
\begin{equation}
H_V=\frac{1}{q!}\sum_{i_1i_2...i_q}i^{q/2}J_{i_1i_2...i_q}\chi_{i_1}\chi_{i_2}...\chi_{i_q}+\frac{1}{2}\sum_{i_{1}i_{2}}iV_{i_{1}i_{2}}\chi_{i_{1}}\chi_{i_{2}}.
\end{equation}
Near the SYK$_q$ fixed point, the SYK$_2$ term is relevant, which governs the low energy physics as a Fermi liquid. Consequently, there is no extensive zero temperature entropy. Below we focus on the $q=4$ case. For small $V/J \ll 1$, the crossover from Fermi liquid (described by SYK$_2$) to the non-Fermi liquid (described by SYK$_4$) happens at a temperature $T\sim V^2/J$, where the thermodynamical entropy increases rapidly from $\mathcal{S}/N\sim 0$ to $\mathcal{S}/N\sim O(1)$.

The path integral formulation for computing the entropy with this additional random hopping term is given by modifying the action \eqref{action} and thus \eqref{SD}. The additional term in the action is:
\begin{equation}
\Delta S^{(2)}=-\frac{V^2}{4N}\int d\tau d\tau' \left(MG_A(\tau,\tau')+(N-M)G_B(\tau,\tau')\right)^2,
\end{equation}
and consequently
\begin{equation}
\Sigma_A=\Sigma_B=J^2(\lambda G_A+(1-\lambda)G_B)^{q-1}+V^2(\lambda G_A+(1-\lambda)G_B).
\end{equation}

\begin{figure}[t]
  \center
  \includegraphics[width=1\columnwidth]{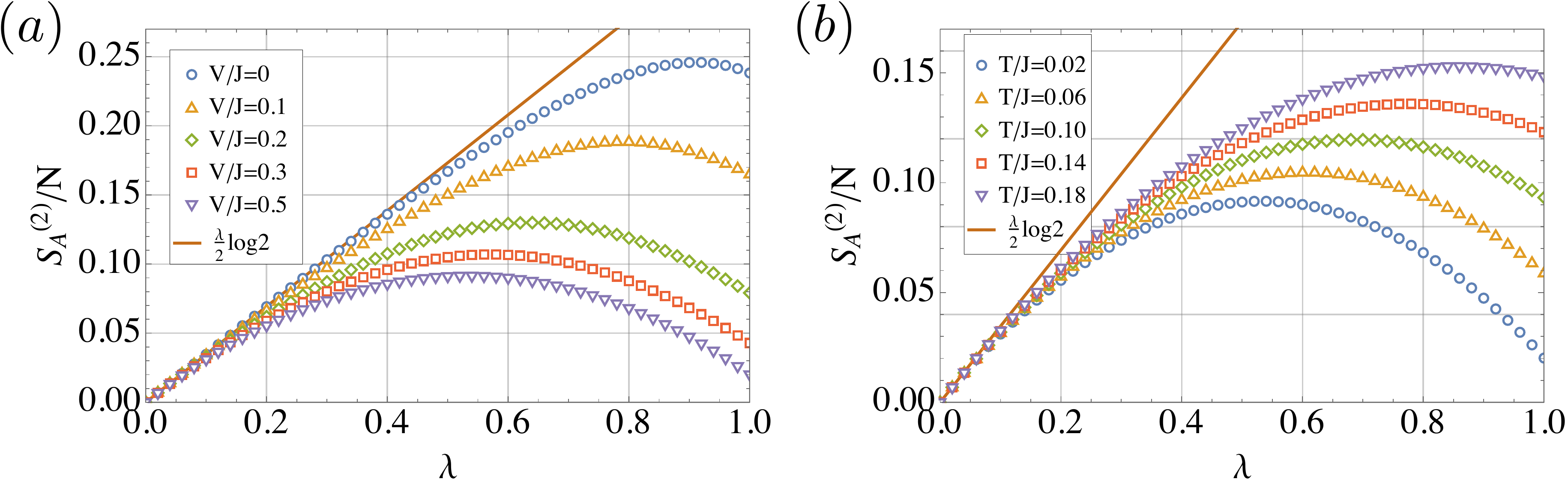}
  \caption{(a). Subsystem entropy $\mathcal{S}_A^{(2)}/N$ for different $V/J$ with $\beta J=50$ and $q=4$. (b). Subsystem entropy $\mathcal{S}_A^{(2)}/N$ for different temperature $T/J$ with $q=4$ and $V/J=0.5$. In both figures we have also plotted $\lambda \log 2/2$ for reference.}\label{fig3}
 \end{figure}

To study this crossover from the entropy perspective, we could either consider tunning $V/J$ with fixed temperature $T/J$ (Shown in Figure \ref{fig3} (a)) or tunning $T/J$ with fixed temperature $V/J$ (Shown in Figure \ref{fig3} (b)). We find that the slopes for $\mathcal{S}_A^{(2)}$ at small $\lambda$ collapse to $\lambda \log 2/2$, since the argument in the last section also works here. In the low-temperature Fermi liquid phase dominated by SYK$_2$, the entropy of the full system is almost zero and $\mathcal{S}_A^{(2)}$ is almost symmetric around $\lambda=1/2$, while for the non-Femri liquid phase the entropy of the total system is always large.

\subsection{Complex SYK model with chemical potential}
Here we consider another generalization with charge conservation called the complex SYK model. Instead of Majorana fermions, we consider complex fermions with annihilation (creation) operator  $c_i$ ($c_i^\dagger$) satisfying $\{c_i,c^\dagger_j\}=\delta_{ij}$ with $i=1,2,...,N$. The Hamiltonian reads
\begin{equation}
H_c=\frac{1}{(q/2)!(q/2)!}\sum_{i_1i_2...i_{q/2}}\sum_{j_1j_2...j_{q/2}}J_{i_1i_2...i_{q/2};j_1j_2...j_{q/2}}c^\dagger_{i_1}c^\dagger_{i_2}...c^\dagger_{i_{q/2}}c_{j_1}c_{j_2}...c_{j_{q/2}}.
\end{equation}
The total charge $Q=\sum_ic_i^\dagger c_i$ is conserved $[Q,H_c]=0$. One could consider an additional chemical potential term by adding $H_c-\mu Q$, which changes the filling $\mathcal{Q}=\left<Q\right>/N$. The $\mathcal{Q}$ dependence of the zero temperature entropy has been worked out in \cite{davison2017thermoelectric,gu2019notes}. When the chemical potential becomes sufficiently large in the low temperature limit, a first order transition is found numerically, where charge $\mathcal{Q}$ jumps from finite to zero \cite{azeyanagi2018phase}. Here we again focus on the $q=4$ case. 

For the complex fermion case, we define the subsystem $A$ by containing $M$ complex fermion modes. With the chemical potential $\mu$, the $G-\Sigma$ action is now written as 
\begin{equation}
\begin{aligned}
S^{(2)}=&-M\log \underset{A}{\det} (\partial_\tau-\mu-\Sigma_A)-(N-M)\log \underset{B}{\det} (\partial_\tau-\mu-\Sigma_B) \\
&-M\int d\tau d\tau'G_A(\tau',\tau)\Sigma_A(\tau,\tau')-(N-M)\int d\tau d\tau'G_B(\tau',\tau)\Sigma_B(\tau,\tau')\\
&-\frac{J^2}{4N^3}\int d\tau d\tau' \tilde{G}(\tau,\tau')^{2}\tilde{G}(\tau',\tau)^2,
\end{aligned}
\end{equation}
with $\tilde{G}(\tau,\tau')=\lambda G_A(\tau,\tau')+(1-\lambda)G_B(\tau,\tau')$. The saddle point equation becomes
\begin{equation}
\begin{aligned}
&G_A=(\partial_\tau-\mu-\Sigma_A)^{-1}_A,\ \ \ \ \ \ G_B=(\partial_\tau-\mu-\Sigma_B)^{-1}_B,\\
&\Sigma_A(\tau,\tau')=\Sigma_B(\tau,\tau')=-J^2\tilde{G}(\tau,\tau')^{2}\tilde{G}(\tau',\tau).
\end{aligned}
\end{equation}
For the particle-hole symmetric case with $\mu=0$, this is proportional to the Majorana case \eqref{action} by a factor of $2$, since in this case $G_s(\tau,\tau')=-G_s(\tau',\tau)$.

\begin{figure}[t]
  \center
  \includegraphics[width=1\columnwidth]{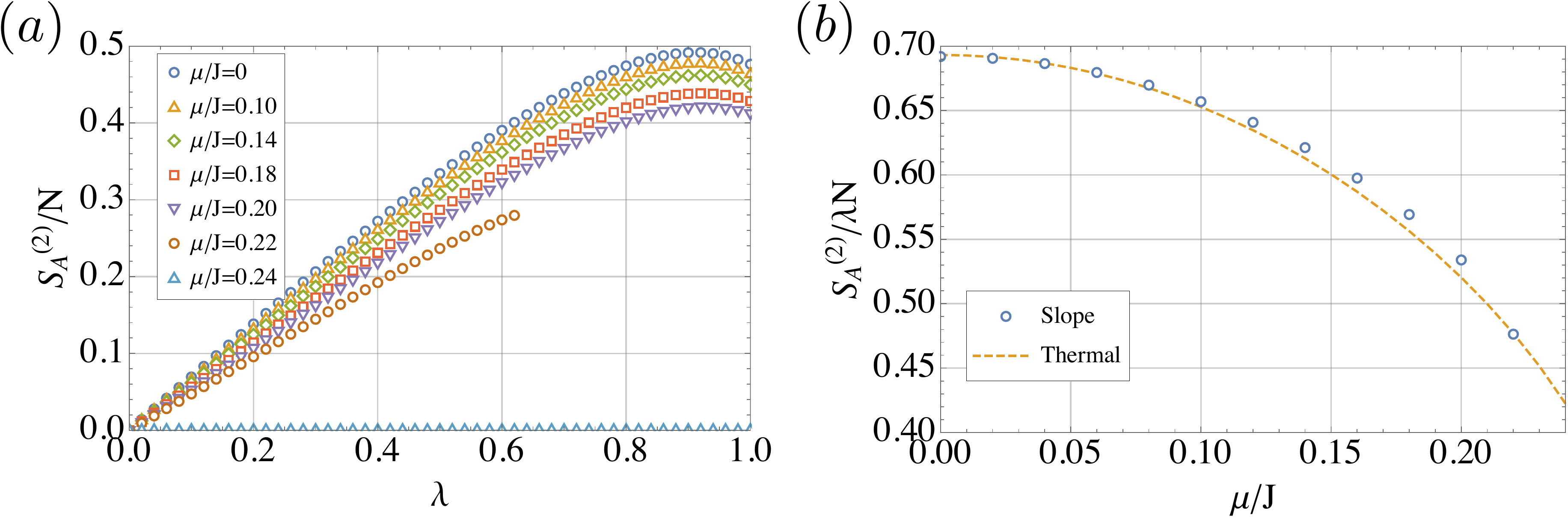}
  \caption{(a). Subsystem entropy $\mathcal{S}_A^{(2)}/N$ for different $\mu/J$ with $\beta J=50$ and $q=4$. (b).The slope of $\mathcal{S}_A^{(2)}/N$ at small $\lambda$ for different $\mu/J$ with $\beta J=50$ and $q=4$.}\label{fig4}
 \end{figure}

For a small subsystem $A$, we expect it is maximally entangled under the density constrain due to its strong correlation with the rest of the system. This means that for a single qubit, in particle number basis, the density matrix should be $\rho_A=\mathcal{Q}|1\rangle\langle1|+(1-\mathcal{Q})|0\rangle\langle0|$. This leads to a R\'enyi entropy $-\log(\mathcal{Q}^2+(1-\mathcal{Q})^2)$, smaller than $\log2$ away from half filling $\mathcal{Q}=1/2$. The numerical result is shown in Figure \ref{fig4} (a). We find that when $\mu/J$ becomes larger, the slope of $\mathcal{S}_A^{(2)}$ at small $\lambda$ becomes smaller. For larger $\mu/J=0.22$, we find a transition of saddle point solution at $\lambda=0.62$ when tuning $\lambda$. This is the analogy of the first order transition in thermal ensemble. For larger $\mu/J$, the saddle point solution converges to the other solution even at $\lambda=0$.

In Figure \ref{fig4} (b), we compare the fitting of initial slope to the prediction from $-\log(\mathcal{Q}^2+(1-\mathcal{Q})^2)$. The charge $\mathcal{Q}$ is determined numerically using a thermal ensemble. We find the formula gives a good approximation. The difference should be attributed to the fact that the relation between $\mu$ and $\mathcal{Q}$ may receive corrections from the replicated system.

\section{Conclusion}
\label{sec:conclusion}

 In this work, we use the path integral approach to analytically derive the self-consistent equation governing the entropy of a subsystem for a thermal ensemble of the SYK-like models in the large-$N$ limit. We numerically solve the equation and study the scaling of subsystem R\'enyi entropy. In particular, we focus on the second R\'enyi entropy and we find results of SYK model at extremely low temperature are consistent with exact diagonalization \cite{liu2018quantum} and can be well approximated by thermal entropy with an effective temperature \cite{huang2019eigenstate} when $\lambda\leq 1/2$. We also study the scaling of R\'enyi entropy as we vary the parameters $q$ and temperature.

We further study the generalizations of the SYK model by introducing the quadratic random hopping term or considering the complex fermion version with $U(1)$ symmetry. For the random hopping case, we see the crossover from the low-energy Fermi liquid to the non-Fermi phase at intermediate temperature. For the complex SYK case, we observe  the chemical potential dependence of entropy for $\lambda\rightarrow 0$. We also find  that a first order transition exists when tuning the subsystem size at large chemical potential.

\section*{Acknowledgements}
We thank Leon Balents, Yiming Chen, Tarun Grover, Yingfei Gu and Yichen Huang for helpful discussions. PZ would like to thank his wife Ning Sun for helping him improve the figures drawing. 

\paragraph{Funding information}
PZ acknowledges support from the Walter Burke Institute for Theoretical Physics at Caltech. 
CL was supported by the DOE, Office of Science, Basic Energy Sciences under award no. DE-FG02-08ER46524.

\begin{appendix}

\section{Subsystem R\'enyi entropy of the SYK$_2$ model}

In this Appendix we derive an analytic form for the R\'enyi entropy of the complex SYK${}_2$ models. We will only present zero temperature result; the more interesting problem of deriving the (possibly) analytical form of R\'enyi entropy of thermal SYK2 ensembles is left to future work. The crucial part is to obtain the eigenvalue distribution of $\rho^{\alpha}_A$, i.e. the $\alpha$-th power of the eigenvalue $\rho_A$, at finite temperatures. The eigenvalue distribution of $\rho_A$ at zero temperature has been derived in Ref.~\cite{liu2018quantum}, which is
\begin{equation}\label{ES1}
f(x,\kappa,\lambda)=
\frac{1}{2\pi\lambda}\frac{\sqrt{(\lambda_+-x)(x-\lambda_-)}}{x(1-x)} 1_{[\lambda_-,\lambda_+]}+
\Theta(\lambda-\kappa)\delta(x)\left(1- \frac{\kappa}{\lambda}\right), \quad \lambda\in (0,1/2],
\end{equation}
where 
\begin{equation}\label{lpm}
\lambda_\pm = \left( \sqrt{\kappa(1-\lambda)} \pm \sqrt{ \lambda(1-\kappa)}\right)^2,
\end{equation}
$\kappa$ denotes filling fraction and $\lambda$ denotes subsystem-system ratio. The Majorana case is recovered by imposing particle-hole symmetry with $\kappa=1/2$. We then compute the generic R\'enyi entropy by
\begin{equation}
\mathcal{S}_A^{(\alpha)}(\kappa,\lambda)
=\frac{1}{1-\alpha} \int f(x) \ln \left(x^\alpha+(1-x)^\alpha\right)dx.
\end{equation}
We notice that for generic $\alpha$ the integral is hard to solve. We therefore restrict ourselves to the case where $n$ is positive integer, and we have
\begin{equation}
\mathcal{S}_A^{(n)}(\kappa,\lambda)=\frac{1}{1-n} \sum\limits_{j=0}^{n-1} \int f(x) \ln \left(x-\zeta_j (1-x)\right) dx,
\end{equation}
where we defined $\zeta_j = e^{\frac{i \pi}{n}(2j+1)}$, $j=0,1,...,n-1$. We just have to calculate
\begin{equation}\label{Ij}
I_j = \int^b_a \frac{\sqrt{(b-x)(x-a)}}{x(1-x)} \ln\left(x - d_j\right)dx,
\qquad d_j \equiv \frac{\zeta_j}{1+\zeta_j}.
\end{equation}
Note this implies that $\zeta_j\neq -1$, meaning that $n$ must be even. If $n$ is odd then there is one $j$ s.t. $\zeta_j = -1$, and we have to treat this term separately. For simplicity we restrict ourselves to even $n$ cases in the following derivation. Using
\begin{equation}\label{xabcdx}
\begin{aligned}
&\int\frac{\sqrt{(b-x)(x-a)}}{x(c-x)}dx \\
&=-2 \arctan\sqrt{\frac{b-x}{x-a}}+2 \frac{\sqrt{ab}}{c} \arctan \sqrt{\frac{a}{b}}\sqrt{\frac{b-x}{x-a}} + 2\frac{\sqrt{(c-a)(c-b)}}{c}\arctan \sqrt{\frac{c-a}{c-b}}\sqrt{\frac{b-x}{x-a}},
\end{aligned}
\end{equation}
where $0<a<x<b<c$ is assumed, we can then integrate Eq.~\eqref{Ij} by parts
\begin{equation}
\begin{aligned}
I_j
=&\pi\left(1-\frac{\sqrt{ab}}{c}-\frac{\sqrt{(c-a)(c-b)}}{c}\right)\ln(a-d_j)
+2I_{1,j}(1)-2 \frac{\sqrt{ab}}{c}I_{1,j}\left(\sqrt{\frac{a}{b}}\right)\\
&-2 \frac{\sqrt{(c-a)(c-b)}}{c}I_{1,j}\left(\sqrt{\frac{c-a}{c-b}}\right),
\end{aligned}
\end{equation}
where
\begin{equation}
I_{1,j}(\eta) = \int \arctan \eta \sqrt{\frac{b-x}{x-a}} \frac{1}{x-d_j} dx
=\pi \ln \frac{1+\sqrt{\frac{b-d_j}{a-d_j}}\eta}{1+\eta}.
\end{equation}
We then have
\begin{equation}
\begin{aligned}
&\mathcal{S}_A^{(n)}(\kappa,\lambda)\\
&=\frac{1}{1-n} \frac{1}{2\pi\lambda}\left[\sum\limits_{j=0}^{n-1}\ln(1+\zeta_j)\int^{\lambda_+}_{\lambda_-} \frac{\sqrt{(\lambda_+-x)(x-\lambda_-)}}{x(1-x)} dx
+\sum\limits_{j=0}^{n-1}I_j\right]\\
&=\frac{1}{1-n} \frac{1}{2 \lambda}\left[\left(1-\sqrt{\lambda_-\lambda_+} - \sqrt{(1-\lambda_-)(1-\lambda_+)}\right)\ln 2+2\ln \frac{\prod\limits_{j=0}^{n-1}\left(\sqrt{\lambda_--d_j}+ \sqrt{\lambda_+-d_j}\right)}{2^n}\right.\\
&\qquad\qquad\qquad\quad\left.-2\sqrt{\lambda_-\lambda_+} \ln \frac{\prod\limits_{j=0}^{n-1}\left(\sqrt{\lambda_+}\sqrt{\lambda_--d_j}+ \sqrt{\lambda_-}\sqrt{\lambda_+-d_j}\right)}{(\sqrt{\lambda_-}+\sqrt{\lambda_+})^n}\right.\\
&\qquad\qquad\qquad\quad\left.-2\sqrt{(1-\lambda_-)(1-\lambda_+)}\ln \frac{\prod\limits_{j=0}^{n-1}\left(\sqrt{1-\lambda_+}\sqrt{\lambda_--d_j}+ \sqrt{1-\lambda_-}\sqrt{\lambda_+-d_j}\right)}{(\sqrt{1-\lambda_-}+\sqrt{1-\lambda_+})^n}
\right],
\end{aligned}
\end{equation}
where
$$n=2,4,6,..., \qquad d_j = \frac{\zeta_j}{1+\zeta_j},\qquad \zeta_j = e^{\frac{i \pi}{n} (2j+1)},\quad j = 0,1,...,n-1.$$
Let us specify the branch cut for the squareroot: $\sqrt{z}$ is defined where $\mathrm{Arg}z\in (-\pi,\pi]$, i.e. the branch cut is along the negative $x$ axis. This way we make sure that the term containing $j$ and $j' = n-1-j$ pair up to give real values. We see that the general form is complicated, and the way to analytic continue to other $\alpha$ values is not obvious. 

We now look at several special cases. In the case with $\lambda = \kappa = 1/2$,  we have
\begin{equation}
\mathcal{S}_A^{(n)}(1/2,1/2)
=\frac{1}{1-n}\left(\ln 2+2 \ln \prod\limits_{j=0}^{n-1}\left(\sqrt{-d_j}+\sqrt{1-d_j}\right) - 2n \ln 2\right).
\end{equation}
Plug in $n= 2m$, $d_j = \frac{\zeta_j}{1+\zeta_j}$ and $\zeta_j = e^{\frac{i \pi}{2m}(2j+1)}$, $\zeta_{2m-1-j} = e^{\frac{i \pi}{2m}(4m-2j-1)} = \bar{\zeta}_j$, then
\begin{equation}
\begin{aligned}
\mathcal{S}_A^{(2m)}(1/2,1/2)
&=
\frac{1}{1-n}\left[(1-2n)\ln 2 + 2 \ln \prod\limits_{j=0}^{m-1}
\frac{2+2\cos\left(\frac{\pi}{4m}(2j+1)-\frac{\pi}{2}\right)}{\sqrt{2\left(1+\cos \frac{\pi}{2m}(2j+1)\right)}}\right]\\
&=
\frac{1}{1-n}\left[-2n\ln 2 + 2 \ln B_{n}(1)\right],
\end{aligned}
\end{equation}
where $B_n$ denotes the Normalized Butterworth Polynomials
\begin{equation}
B_{2m}(s) \equiv \prod\limits_{k=0}^{m-1}\left[s^2-2s \cos \left(\frac{\pi}{4m}(2k+1) + \frac{\pi}{2}\right)+1\right].
\end{equation}
As a check: when $n=2$, $B_2(1) = 2+\sqrt{2}$, we have
\begin{equation}
\mathcal{S}_A^{(2)}(1/2,1/2) =-(-4 \ln 2 + 2 \ln (2+\sqrt{2})) = 3\ln 2 - 2 \ln (1+\sqrt{2}).
\end{equation}

Another interesting case is the R\'enyi entropy at $n\rightarrow +\infty$. Using
\begin{equation}
2\ln B_n(1) = 2\ln \prod_{k=0}^{m-1} 2\left(1+ \sin\frac{\pi}{4m}(2k+1)\right)\rightarrow n \ln 2 + \frac{4m}{\pi} \int^{\pi/2}_0 \ln (1+ \sin x) dx=\frac{4C}{\pi} n,
\end{equation}
where $C = 0.91596559...$ is the Catalan constant, we have
\begin{equation}
\mathcal{S}_A^{(\infty)}(1/2,1/2)
=\lim\limits_{n\rightarrow \infty}
\frac{1}{1-n}\left(-2n \ln 2+\frac{4C}{\pi}n\right)
=2 \ln 2 - \frac{4C}{\pi} = 0.220050745.
\end{equation}

The value $\mathcal{S}_A^{(\alpha)}(\kappa=1/2,\lambda=1/2)$ for arbitrary real number $\alpha$ in fact can be obtained:
\begin{equation}
\begin{aligned}
\mathcal{S}_A^{(\alpha)}(1/2,1/2)
&=\frac{1}{1-\alpha}\frac{1}{\pi}\int^1_0 \ln(x^\alpha+(1-x)^\alpha) \frac{1}{\sqrt{x(1-x)}}dx\\
&=\frac{1}{1-\alpha}\frac{2}{\pi} \int^1_{1/2} \left[\ln (x^\alpha)+\ln\left(1+\left(\frac{1-x}{x}\right)^\alpha\right)\right] \frac{1}{\sqrt{x(1-x)}}dx\\
&=\frac{2}{1-\alpha}\frac{\alpha}{\pi}\left(2C-\pi \ln 2\right)+\frac{2}{1-\alpha}\frac{1}{\pi} I(\alpha),
\end{aligned}
\end{equation}
where we defined
\begin{equation}
I(\alpha)
= 2 \int^1_0 \frac{\ln (1+p^{2\alpha})}{1+p^2} dp
\end{equation}
with $ p = \sqrt{\frac{1-x}{x}}$. We can obtain, for example, $\mathcal{S}_A^{(0)}(1/2,1/2) = \ln 2=0.693147$, $\mathcal{S}_A^{(1/2)}(1/2,1/2) = \frac{4C}{\pi}-\ln 2=0.473096$, and the von Neumann value $\mathcal{S}_A^{(1)}(1/2,1/2)= 2 \ln 2-1 = 0.386294$.

\end{appendix}



\bibliography{ref.bib}

\nolinenumbers

\end{document}